\documentclass[doublecol]{epl2}

\usepackage{graphicx}

\def\<{\langle}
\def\>{\rangle}
\def\({\left(}
\def\){\right)}
\def\[{\left[}
\def\]{\right]}

\def \ve {\varepsilon}

\title{Glass-like Thermal-Transport in Symmetry-Broken Clathrates}

\author{Eiji Kaneshita\inst{1,2}\footnote{E-mail: \email{knsht@yukawa.kyoto-u.ac.jp}}
\and Tsuneyoshi Nakayama\inst{3,4}
}
\shortauthor{E. Kaneshita \etal}

\institute{
  \inst{1} Advanced Photon Source, Argonne National Laboratory, 9700 South Cass Avenue, Argonne, IL 60439, USA \\
  \inst{2} Yukawa Institute for Theoretical Physics, Kyoto University, Kyoto 606-8502, Japan \\
  \inst{3} Materials Science Division, Argonne National Laboratory, 9700 South Cass Avenue, Argonne, IL 60439, USA \\
  \inst{4} Toyota Physical and Chemical Research Institute, Nagakute, Aichi 480-1192, Japan
}

\pacs{61.72.Bb}{Theories and models of crystal defects}
\pacs{66.35.+a}{Quantum tunneling of defects}
\pacs{61.43.-j}{Disordered solids}

\abstract{
We present the quantitative interpretation for the glass-like behavior of thermal conductivities $\kappa(T)$ for type-I clathrate compounds involving off-centered guest ions.
It is shown that the dipole-dipole interaction generated in cage/guest-ion systems is crucial to reproduce the characteristics of thermal conductivities for these symmetry-broken clathrates.
The above scenario also explains well the difference of $\kappa(T)$ between the $p$-type and the $n$-type $\beta$-BGS found recently by K. Suekuni \textit{et~al.} [\textit{Phys. Rev. B}, \textbf{77}~(2008)~235119.]
}

\begin{document}

\maketitle

\section{Introduction}
The ``phonon-glass electron-crystal" concept, proposed over a decade ago~\cite{Sla95} in order to explore thermoelectric materials with the high figure of merits $Z$~\footnote{The dimensionless parameter $ZT$ is often used to evaluate the thermoelectric efficiency at temperature $T$.
Here $Z=S^2/\rho(\kappa_p+\kappa_e)$ with $\rho$ the electrical resistivity, $\kappa_e$ ($\kappa_p$) the electric (phonon) thermal conductivity, and $S$ the thermopower or Seebeck coefficient, respectively.
The low  $\kappa_p$ is crucial to increase the efficiency of thermoelectric conversion since $\kappa_e$ is combined with $\rho$.}, has revived interest in thermal transports in  solids.
Certain classes of thermoelectric materials with low thermal conductivities have been searched, leading to the discovery  of type-I clathrate compounds with the formula II$_8$III$_{16}$IV$_{30}$~\cite{Nol98}, where the group III and IV elements constitute framework atoms of cages and the group II elements are guest ions in the cages.
Subsequent work~\cite{Coh99, Nol00-1, Nol00-2, Sal01, Ben04, Avi06, Sue07, Sue08, Avi08} has experimentally confirmed that the thermal conductivities $\kappa(T)$ of these compounds are not only drastically suppressed but also display characteristics identical to those of network glasses~\cite{Zel71}.

Avila \textit{et~al.}~\cite{Avi08} have achieved the lowest recorded $\kappa(T)$ among clathrate compounds in $\beta$-Ba$_8$Ga$_{16}$Sn$_{30}$ ($\beta$-BGS).
In addition, Suekuni \textit{et~al.}~\cite{Sue08} have recently discovered the clear difference of $\kappa(T)$ between the $p$-type and the $n$-type of $\beta$-BGS.
Actually, they have found that $\kappa_n(T)$ of the $n$-type is smaller than that for $\kappa_p(T)$ of the $p$-type below a few K, while the reverse tendency in the regime above a few K.

\begin{figure}[htbp]
\begin{center}
\includegraphics[width = 0.6\linewidth]{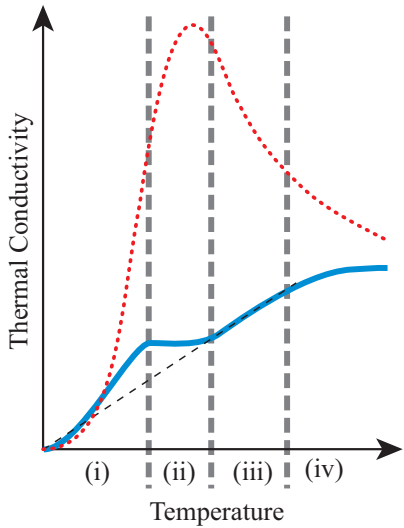}
\caption{(Color online) Scheme of thermal conductivities $\kappa(T)$ of type-I clathrates. The solid line shows the glass-like temperature dependence of $\kappa(T)$ for symmetry-broken clathrate with off-centered guest ions.
The dotted line gives the crystallinelike behaviors of symmetric type-I clathrate without off-centered guest ions.
In the low-temperature region $T<1$~K~(i), the former behaves as $T^2$, while the latter does as $T^3$.
The plateau temperature region appears in the former case at $\sim10$~K.
The thin dashed line depicts the $T$-linear behavior in the temperature region above 100~K~(iii), which approaches zero as $T\rightarrow0$.
From 100~K, $\kappa(T)$ curls over.
} \label{fig:K-T}
\end{center}
\end{figure}

It is noteworthy that thermal conductivities almost identical to those of glasses are observed in the clathrates \textit{with} off-centered guest ions (symmetry-broken clathrates) but \textit{not} in those \textit{without} off-centered guest ions (symmetric clathrates) with translational invariance~\cite{ Nol98, Coh99, Nol00-1, Nol00-2, Sal01, Ben04, Avi06, Sue07, Sue08, Avi08}.
We illustrate a scheme of the characteristics of $\kappa(T)$ in Fig.~\ref{fig:K-T} for both of the clathrates with and without off-centered guest ions.
Figure 1 shows that $\kappa(T)$ varies as $T^{\alpha}$~($\alpha\approx~2)$ at temperatures below a few K.
Above a few K, there appears a plateau in $\kappa(T)$, and it lasts up to 10~K.
Above 10~K, $\kappa(T)$ rises again $T$-linearly.
From 100~K, $\kappa(T)$ curls over.
These are almost identical to those of general glasses with topological disorder~\cite{Nak02}, while $\kappa(T)$ of the symmetric clathrates, depicted by a dotted line in Fig.~\ref{fig:K-T}, shows identical thermal conductivities to those of crystals with translational symmetry.

In this connection, we should mention that orientationally disordered crystallines, so-called orientational glasses, show all of the universal properties of structural glasses~\cite{Loi83, DeY86}.
Grannan~\textit{et al.}~\cite{Gra90} and Randeria and Sethna~\cite{Ran88} have theoretically investigated the role of long-range interaction between elastic dipoles \textit{randomly} distributed in space.
Although their systems are related with the present work, electric dipoles in our systems are \textit{regularly} distributed and the situation is clearer.
Recently, Parshin~\textit{et al.}~\cite{Par07} have discussed the vibrational instability of weakly interacting harmonic modes controlled by the 4th-order anharmonicity of the potential function.
They have demonstrated that the instability produces both of features observed in glasses:  the so-called boson peak and two-level tunneling systems.
Our work takes into account the explicit interaction between electric dipoles in nearby cages for symmetry-broken clathrate systems rather than the harmonic coupling between the oscillators.

In this Letter, we formulate the thermal conductivity of the symmetry-broken clathrates in terms of the interacting-dipole picture and provide a theoretical interpretation on why the symmetry-broken clathrates show the characteristics of $\kappa(T)$ almost identical to those of network glasses, by focusing our attention on the recent experiments for the prototype type-I clathrate, $\beta$-BGS~\cite{Avi06, Sue07, Sue08, Avi08}.
Our concepts are general and not only applicable to $\beta$-BGS but also to other types of clathrates with off-centered guest ions.

\section{Thermal conductivities below a few K}
The positional symmetry of the guest ions in type-I clathrates is broken by increasing the size of the cages, and the guest ions take off-center positions at $r_0$ apart from the center of the cage~\cite{Coh99, Nol00-1, Nol00-2, Sal01,Avi06, Sue07, Sue08, Avi08}.
A typical distortion $r_0=0.43$~{\AA} in $\beta$-BGS has been determined from diffraction experiments~\cite{Avi06, Avi08}.
This value is about 7.2\% of the distance between the neighboring 14-hedron cages ($a/2=5.84$~{\AA}, where $a$ is the lattice constant).

\begin{figure}[htbp]
\begin{center}
\includegraphics[width = 0.9\linewidth]{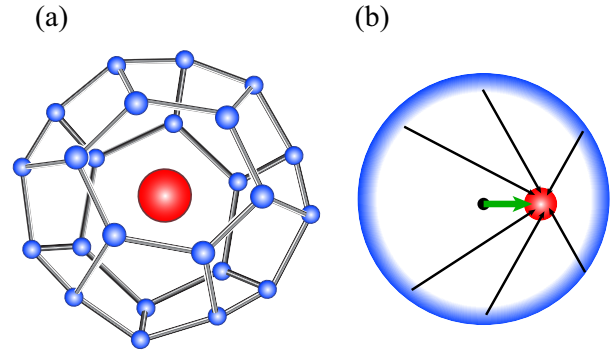}
\caption{(Color online) (a) A 14-hedron cage consisting of anions and a guest cation.
(b) The cage (outer circle) and the off-centered guest ion compose an effective electric dipole moment (thick arrow), which is the vector sum of each dipole (thin arrow).
} \label{fig:dipole}
\end{center}
\end{figure}

The deviation of a guest ion from the center of the cage in $\beta$-BGS induces a large electric-dipole moment, which is due to the difference of the charge between the Ba$^{2+}$ guest ion and the Ga$^{-}$ ions of the cage (see Fig.~\ref{fig:dipole}) as pointed out in~\cite{Nak08}.
The strength of the electric dipole moment becomes $p\approx 4.1$ Debyes for $\beta$-BGS.
Thus, the effect of the dipolar interaction between guest ions becomes crucial since the coupling energy of the nearest neighbor dipole-dipole interaction is of the order of 60 K~\cite{Nak08}.

Of importance is that the long-range and anisotropic dipole-dipole interaction constitutes many potential minima in the configuration space of multiple dimensions; therefore, the quantum states should be represented in a configuration space.
There should be many stable configurations, and we represent the $i$-th stable configuration by $X_i$ in the configuration space of multiple dimensions.
The stable states should execute the zero-point oscillations (quasi-ground states) around the stable configurations.

\begin{figure}[htbp]
\begin{center}
\includegraphics[width = 0.9\linewidth]{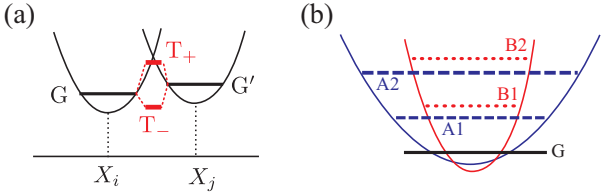}
\caption{(Color online) Scheme of states in configuration space.
The states for different dipole-configurations $X_i$ and $X_j$ are expressed by G ($=|X_i,\{0\}\>$) and G' ($=|X_j,\{0\}\>$).
The hybridization of these states creates two-level tunneling states $\mathrm{T}_\pm$ ($=|(X_i,X_j),\pm\>$).
(b) The excited states $|X_i,\{n_\lambda\}\>$ (labeled as A1, B1, etc.) at a certain configuration $X_i$ is illustrated.  The transition between the excited states A1 and B1 corresponds to a hopping of the quasilocalized oscillation in the real space~(See the text).
} \label{fig:config}
\end{center}
\end{figure}

When two stable states $|X_i,\{0\}\>$ and $|X_j,\{0\}\>$ exist near each other in the configuration space, the two-level tunneling states $|(X_i,X_j),\pm\>$ are constructed from these states, where $+$ ($-$) denotes the higher (lower) energy state.
The conceptional illustration of the tunneling states in the configuration space is shown in Fig.~\ref{fig:config}~(a).
The point is that the tunnelings occur in the configuration space of multiple dimensions but not in a single cage.

The thermal conductivity is expressed as
\begin{eqnarray}
\kappa(T)&=&\sum_{\epsilon}\int_0^{\infty} \hbar\omega \frac{\partial n_{\mathrm{BE}}(\omega,T)}{\partial T}
v_\epsilon^2 \tau_\epsilon D_\epsilon(\omega)
\,d \omega,
\label{eq:kappa}
\end{eqnarray}
where $n_{\mathrm{BE}}$ is the Bose-Einstein distribution function, and $v_{\epsilon}$, $\tau_{\epsilon}$, and $D_\epsilon(w)$ are the group velocity $\partial\omega (k)/\partial k$, life-time, and density of states of thermal phonons of the mode $\epsilon$, respectively.
We take
\begin{eqnarray}
 D_\epsilon(\omega)=\left\{
\begin{array}{cl}
\frac{\omega^2}{2\pi^2v_{\epsilon}^3}& \mbox{for $\omega\leq\tilde{\omega}_{\epsilon}$}\\
0 & \mbox{for $\omega>\tilde{\omega}_{\epsilon}$}
\end{array}\right.,
\end{eqnarray}
where $\tilde{\omega}_{\epsilon}$ is the upper limit of the extended acoustic-phonons with linear dispersion.
For $v_{\epsilon}$, we take the average $v_{\mathrm{s}}=2400$~$\mathrm{m\,s^{-1}}$ obtained from the observed velocities for $\beta$-BGS~\cite{Sue08}:
$v_{\mathrm{l}}=3369$~$\mathrm{m\,s^{-1}}$ for the longitudinal phonons and $v_{\mathrm{t}}=1907$~$\mathrm{m\,s^{-1}}$ for the transverse phonons.
Note that the Matthiessen's rule $1/\tau_{\epsilon}=\sum_i 1/\tau_{\epsilon, i}$ should hold when there are different scattering processes.

In the low-temperature regime below a few K, only two-level tunneling states $|(X_i,X_j),\pm\>$ become relevant.
The transition between $|(X_i,X_j),-\>$ and $|(X_i,X_j),+\>$ is induced by a resonant absorption or emission of the thermal phonons, and this interaction yields the scattering rate of the form
\begin{eqnarray}
\frac{1}{\tau_{\mathrm{tun}}(\omega)}=\frac{2\pi^{3/2}\ve_r g^2 \omega}{\rho v_{\epsilon}^2}\(\frac{\eta}{p^2}\)
\tanh\(\frac{\hbar\omega}{k_B T}\),
\label{eq:scatrate}
\end{eqnarray}
where $g$ is the deformation coupling constant, and $\rho$ the mass density ($\rho=6010$~$\mathrm{kg\,m^{-3}}$ for $\beta$-BGS).
$\ve_r$ is the dielectric constant, and $\eta$ equals to $n_{\mathrm{tun}} / n_{\mathrm{di}}$, where $n_{\mathrm{tun}}$ and $n_{\mathrm{di}}$ are the numbers of the tunneling states and the actual dipoles per volume, respectively; \textit{i.e.}, $1/\eta$ is equivalent to the averaged number of off-centered guest ions involved in a dipole rearrangement caused by tunneling~\cite{Nak08} and is a function of $p$.
It should be noted that the analysis of the specific heats for $\beta$-BGS provides the average number of off-centered guest ions $1/\eta\simeq$~20~\cite{Nak08}, implying the failure of independent guest ion picture.

By combining the above scattering rate with Eq.~(\ref{eq:kappa}), $\kappa(T)$ of the type-I clathrates becomes
\begin{eqnarray}
\kappa(T) = \frac{3\rho k_B^3 v_s }{4\pi^{3/2}\hbar^2g^2\varepsilon_r (\eta/p^2)}T^2.
\label{eq:kappa-T2}
\end{eqnarray}
The convenient form with the specific heat $C_{\mathrm{tun}}$ is obtained as
\begin{eqnarray}
\kappa(T)
&=&\(\frac{\pi N_A k_B^3}{4\hbar^2}\)
\(\frac{\rho a^3v_s}{C_{\mathrm{tun}}/T}\)\(\frac{T}{g/k_B }\)^2
\label{eq:kappa-C}\\
&=& 86000\(\frac{T}{g/k_B}\)^2\mathrm{[W\,K^{-1}\,m^{-1}]}.
\end{eqnarray}
where
\begin{eqnarray}
C_{\mathrm{tun}}=\frac{\pi^{5/2} \ve_r a^3 N_Ak_B^2}{3}\(\frac{\eta}{p^2}\)T.
\label{eq:C3}
\end{eqnarray}
From the experimental data of $\kappa (T)\simeq~0.02T^2$ $\mathrm{W\,K^{-1}\,m^{-1}}$ and $C(T)\simeq~30T$ $\mathrm{mJ\,mol^{-1}\,K^{-1}}$ below a few K for symmetry-broken $\beta$-BGS~\cite{Sue08}, the deformation coupling $g$ is estimated as $g\simeq0.2$~eV ($g/k_B\simeq2100$~K).
This is a reasonable value because the deformation coupling constants in glasses are in the range of 0.1-1.0~eV~\cite{And72,Phi72}.

It is worthwhile to consider the effect of the strength change of electric dipole moment $p$ on $\kappa(T)$.
Suekuni \textit{et al.}~\cite{Sue08} have observed the difference of thermal conductivities between the $n-$type and the $p-$type $\beta$-BGS.
The observed thermal conductivities below a few K of the $n$-type $\beta$-BGS are a few percent smaller than those of the $p$-type~\cite{Sue08}.
This difference of $\kappa(T)$ should arise from the difference of the dipole moments:
The deviation of the guest ions is different between the $n$-type ($r_0=0.434$) and the $p$-type (0.439~{\AA}).
This effect appears in Eq.~(\ref{eq:kappa-T2}) as the factor $\eta/p^2$.
(Note that $\eta$ should also depend on $p$.)

Since the dipolar interaction is proportional to $p^2/r^3$, where $r$ is the distance between the dipoles, the effective radius $r_{e}$ of the dipolar interaction is proportional to $p^{\frac{2}{3}}$.
Therefore, the average number of dipoles involved in the tunneling, $1/\eta$, should be proportional to $p^2$, then it follows that $p^2/\eta \propto p^4$ or
\begin{eqnarray}
\eta_n = \eta_p\(\frac{p_p}{p_n}\)^2.
\label{eq:eta}
\end{eqnarray}

Thus, the difference of the dipoles, $p_p>p_n$, leads to the relation $\kappa_p>\kappa_n$ below a few K, while the observed specific heats behave the other way, $C_{\mathrm{tun},p}(T) < C_{\mathrm{tun},n}(T)$~\cite{Sue08}.
Both characteristics are consistent with our formulation in Eq.~(\ref{eq:kappa-C}), where $C_{\mathrm{tun}}$ appears in the denominator.
Our formulation reproduces the different tendency between the specific heats and the thermal conductivities for the $n$-type and the $p$-type $\beta$-BGS.

It should be noted that the above formulations are valid only when the system can be well-described by the interacting-dipole picture.
When $p$ is very small, Eqs.~(\ref{eq:C3}) and (\ref{eq:scatrate})-(\ref{eq:kappa-C}) are not valid.
In the above formulation, we assumed the energies of the tunneling states are uniformly distributed, which should be true only for the states distributed in a wide energy range.
For a small $p$, however, the states are distributed in a narrow energy range or do not exist.

For such a small $p$, the scattering rate by the tunneling states must become small as the dipolar feature weakens, \textit{i.e.}, $\tau_{\mathrm{tun}}^{-1}\rightarrow0$ as $p\rightarrow0$.
Then, the total scattering rate of thermal phonons is given by $\tau^{-1}=\tau_{\mathrm{tun}}^{-1}+\tau_{\mathrm{bou}}^{-1}$, where $\tau_{\mathrm{bou}}^{-1}$ ($=\mathrm{const.}$) is the boundary scattering of the system and constant.
This yields a constant $\tau^{-1}$, leading the same $T^3$-dependence of $\kappa(T)$ at low temperatures as observed in symmetric clathrates.
Thus, $\kappa(T)$ becomes more crystalline-like as $p$ decreases.
The temperature dependence changes from $T^2$ to $T^3$ when $p$ gets close to zero, as observed experimentally~\cite{Sue08}.

\section{Plateau of $\boldsymbol{\kappa(T)}$ above a few K}
Observed thermal conductivities $\kappa(T)$ for $\beta$-BGS \cite{Sue08} show the plateau behavior above $\sim3$~K and increase again T-linearly above $\sim10$~K.
In addition, the experiments~\cite{Sue08} show the reverse tendency \mbox{$\kappa_p<\kappa_n$} at the plateau, compared with the case of \mbox{$\kappa_p>\kappa_n$} below $\sim3$~K.
The Rayleigh scattering is negligible owing to the irrelevance of mass-density fluctuation in the length scale of excited-phonon wavelengths below a few K.

These observations indicate that other types of oscillating states become relevant:  for example, angular oscillation modes of dipoles as in the case of orientational glasses~\cite{DeY86,Loi83,Gra90,Ran88}.
The eigenfrequency of angular oscillation mode, $\omega_\theta$, should exhibit the moment-of-inertia effect \mbox{$\omega_\theta\propto I^{-1/2}$} with the definition $I=m_{Ba}r_0^2$, whose effect has been clearly demonstrated for orientational glasses~\cite{Tal02}.
For $\beta$-BGS, the moments of inertia provide the inequality $I_p>I_n$, which leads the inequality $\omega_{\theta, p}<\omega_{\theta, n}$ for $p$-type and $n$-type $\beta$-BGS.
The coupling of acoustic phonons with angular oscillation modes composes the flattened and anti-crossing dispersion curve.

The expression of $\kappa(T)$ in Eq.~(\ref{eq:kappa}) suggests that, if the scattering rate of thermal phonon $\tau^{-1}$ is independent of $T$, $\kappa(T)$ takes the same temperature dependence as the Dulong-Petit limit of the Debye specific heat in the high-temperature region, \textit{i.e.}, $\kappa(T)=\mathrm{const}$.
As demonstrated in Ref.~\cite{Nak98}, the dominant phonons at
$k_BT\geq\hbar\tilde{\omega}_{\epsilon}$, namely, the
coupled modes behave as strongly localized modes satisfying the condition $\omega_\theta\tau(\omega_\theta)\approx$ 1.
Under the situation $k_BT>\hbar\tilde{\omega}_{\theta}$, $\kappa(T)$ behaves $\propto \omega_{\theta}^2$ from Eq.~(\ref{eq:kappa}). Hence, we obtain  the plateau feature in addition to the type-dependence, $\kappa_p(T)<\kappa_n(T)$.

In this connection, we should mention the experimentally observed broad peak in specific heat $C(T)/T^3$ at around 5~K for $\beta$-BGS~\cite{Avi06, Sue07, Sue08, Avi08}.
Similar broad peaks are observed in all glasses (for network glasses, see, \textit{e.g.}, Ref.~\cite{Nak02}) and are called the boson peak.
The boson peak in the specific heat refers to an excess contribution of an excitation over the usual Debye density of states~\cite{Nak02}.
It is natural to expect that the boson-peak-like spectra for symmetry-broken clathrates should also be observed in spectroscopic measurements.
Such peaks have been surely found in Raman scattering measurements for $\beta$-XGG~(X=Ba, Sr, Eu)~\cite{Tak06, Tak08}.
$\beta$-EGG, a Symmetry-broken clathrate, shows boson-peak-like modes in the Raman scattering spectra with a broad peak around a few 10~$\mathrm{cm^{-1}}$~\cite{Tak06, Tak08}.

In the case of silica glass($v$-SiO$_{2}$), hyper-Raman scattering experiments performed by Helen et al.~\cite{Hel00} have definitely shown the existence of hyper-Raman active modes at slightly lower energies than those of Raman active modes.
Thus, there should exist corresponding additional modes
(angular-oscillation modes) with broad spectra for symmetry-broken clathrates.
 These are weakly overlapped with Raman active modes as in the case of $v$-SiO$_{2}$.
The existence of these modes should be confirmed by hyper-Raman, infrared, or inelastic neutron scattering experiments.

We should mention that the symmetric clathrates, $\textit{i.e.}$, the type-VIII clathrates or type-I clathrates without off-centered guest ions ($\textit{e.~g.}$ $\alpha$-BGS; $\beta$-$\mathrm{Sr_8Ga_{16}Si_{30}}$, \textit{etc.}) show  similar peaks  in the specific heats at higher temperatures~\cite{Sue07, Sue08}.
In these symmetric compounds, the excess specific heats have been interpreted in terms of a harmonic oscillator, \textit{i.e.}, the Einstein model~\cite{Sue07, Sue08}, unlike for the symmetry-broken clathrates.

\section{$T$-linear rise above plateau region}
At temperatures above the plateau region, the thermal conductivities follow the relation $\kappa(T)=\gamma T$~\cite{ Coh99, Nol00-1, Nol00-2, Sal01, Ben04, Avi06, Sue07, Sue08, Avi08}.
Actually, the factor $\gamma$ takes $\gamma=0.009$~$\mathrm{W\,K^{-2}\,m^{-1}}$ for the $n$-type and $\gamma=0.007$~$\mathrm{W\,K^{-2}\,m^{-1}}$ for the $p$-type $\beta$-BGS~\cite{Avi06, Sue07, Sue08, Avi08}.
Of importance is that these behave as $\kappa(T)\rightarrow0$ when extrapolating the temperature as $T\rightarrow0$~(see Fig.~\ref{fig:K-T}), although this feature has not been pointed out so far.
It implies that a new \textit{additional} heat-transport channel opens up above the plateau region~\cite{Ale86, Nak99}.
The additional channel is associated with the hopping of the quailocalized oscillations kicked by the thermal phonons as shown below.

\begin{figure}[htbp]
\begin{center}
\includegraphics[width = 0.7\linewidth]{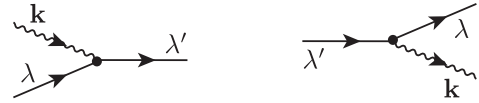}
\caption{The hopping processes of quasilocalized modes.
A thermal phonon (wavy lines) kicks a quasilocalized mode (solid lines).
} \label{fig:hopping}
\end{center}
\end{figure}

We consider an interaction where a thermal phonon with wave vector $\mathbf{k}$ kicks a locally oscillating state, $|X_i,\{n_\lambda\}\>$, into another, $|X_j,\{n_\lambda'\}\>$, which corresponds to the A1$\leftrightarrow$B1 transition in Fig.~\ref{fig:config}~(b).
The interaction Hamiltonian is expressed by
\begin{eqnarray}
H_{\mathrm{hop}}=\Gamma_{\mathrm{eff}} \sum_{\mathbf{k}, \lambda, \lambda'}(A_{\mathbf{k}, \lambda, \lambda'}\, c_{\lambda}^{\dagger} c_{\lambda'} b_{\mathbf{k}}+\mathrm{h.c.}),
\label{H_3}
\end{eqnarray}
where the operators $b^{\dagger}$ and $c^{\dagger}$ ($b$ and $c$) are the creation (annihilation) operators of the thermal phonons and the quasilocalized modes, respectively.
The coefficient $\Gamma_{\mathrm{eff}}$ is the coupling constant of the three phonon processes shown in Fig.~\ref{fig:hopping}.
According to Ref.~\cite{Nak99}, this type of process yields a $T$-linear contribution to the thermal conductivity.
Using the definition $\kappa_{\mathrm{hop}}(T)=\sum_{\lambda}R_{\lambda}^2/\tau_{\mathrm{hop}}(\omega_{\lambda}, T)$ and following a discussion similar to that in Ref.~\cite{Nak99},
the hopping contribution to the thermal transport turns out to be
\begin{eqnarray}
\kappa_{\mathrm{hop}}(T)=\frac{12^2 \Gamma_{\mathrm{eff}}^2 k_B^2T }{\pi^4\rho^3\bar{\ell}^5}\(\frac{1}{v_{\mathrm{l}}^{5}}+\frac{2}{v_{\mathrm{t}}^5}\),
\label{eq:HP}
\end{eqnarray}
where $\rho$ is the mass density, which is $6.01\times10^3$~$\mathrm{kg\,m^{-3}}$ for $\beta$-BGS.
Here, we defined $(4\pi/3)\bar{\ell}^3$ as the average volume for finding a single quasilocalized mode.
Since the contribution of the thermal phonon to the total $\kappa$ is reduced owing to the scattering process by the hopping interaction between the low-energy thermal phonon and the quasilocalized modes (local vibrations bilinearly coupled to the acoustic phonons), we assume $\kappa(T)\approx\kappa_{\mathrm{hop}}(T)$ and obtain the $T$-linear thermal conductivity.

The localization length $\bar{\ell}$ is determined so that $(4\pi/3)\bar{\ell}^3\approx(1/6\eta)a^3$, where 6 is the number of dipoles in the unit cell.
Here, we set $\eta_p\sim0.060$ and $\eta_n\sim0.58\eta_p$ for $\beta$-BGS from Eq.~(\ref{eq:eta}) and the speculation in Ref.~\cite{Nak08}.
From Eq.~(\ref{eq:HP}) and the experimentally observed $\kappa(T)$ ($\propto\gamma T$), the coupling constant $\Gamma_{\mathrm{eff}}$ is estimated as $|\Gamma_{\mathrm{eff}}|\sim 3.1\times10^{13}$ ($2.7\times10^{12}$) for the $n$-type ($p$-type) $\beta$-BGS.
The estimated coupling constants are of the same order as in the case of network glasses~\cite{Yam99}, where the large anharmonicity is the necessary condition for explaining the $T$-linear rise of $\kappa(T)$ above the plateau~\cite{Nak99}.

We mention the numerical simulations of thermal conductivities for amorphous Si~\cite{All89} performed under the following conditions: The system size was $10^3$ atoms with periodic boundary conditions and a harmonic approximation.
The finite size of 16.3~{\AA} prohibits any lower-energy propagating modes below 10~meV, which indicates their results are applicable at higher temperature regime observed for amorphous Si by Cahill et al~\cite{Cah89}.
However, their numerical results do \textit{not} produce the extrapolating behavior, $\kappa(T) \to 0$ at $T\to 0$, experimentally established for glasses and symmetry-broken clathrates.

Propagating acoustic modes at very low-energies do exist in glasses at temperatures and play a role for thermal conductivity above the end of the plateau since they obey the Bose-Einstein distribution.
Our thermal-conductivity mechanism above the plateau highlights the anharmonic interaction between quasilocalized modes and these propagating acoustic phonons incorporating the strong anharmonicity in symmetry-broken clathrates compared with the case of amorphous Si.

\section{Summary}
We have provided the interpretation of the glass-like behavior of the thermal conductivities of symmetry-broken clathrates over a wide temperature range.
Considering the scattering of the thermal phonons by the tunneling sates and the  states coupled with angular orientation modes, in addition to the hopping of these quasilocalized modes, we have reproduced the temperature dependence of $\kappa(T)$ and have obtained the following conclusions.

i) Two-level tunneling states generated in a configuration space under the interacting-dipole picture explain well the observed $T^2$-dependence of $\kappa(T)$ below a few K, and, in addition, the feature \mbox{$\kappa_p>\kappa_n$} between the $p$-type and $n$-type symmetry-broken clathrates below a few K.

ii)  The plateau of $\kappa(T)$ is due to the coupling of acoustic phonons with angular orientation modes of dipoles~\cite{Gra90}, namely, the crossover from extended to quasilocalized phonons (the flattening of dispersion curves of thermal phonons) should occur in the acoustic phonon branch. The observations \mbox{$\kappa_p<\kappa_n$}~\cite{Sue08} at the plateau region arise from the difference of angular oscillation, $\omega_{\theta,p}<\omega_{\theta,n}$.
We have predicted the existence of low-energy modes at around 10~$\mathrm{cm^{-1}}$ for $\beta$-BGS attributing to the angular oscillation modes of dipoles, which are  active to infrared or hyper-Raman scattering.

iii) The observed $\kappa(T)$ above $\sim10$ K for $\beta$-BGS behaves as $\kappa(T)\rightarrow 0$ as $T\rightarrow 0$.
This suggests the onset of a new additional heat transport channel.
We have proposed the new channel associated with the hopping of quasilocalized modes kicked by the thermal phonons, namely, the hopping mechanism of the quasilocalized modes comes up above the plateau temperature region and leads to a heat transport with the $T$-linear dependence of $\kappa(T)$.

\textbf{Acknowledgements} --
This work was supported by the U.S. DOE, Office of Science, Office of Basic Energy Sciences, under Contract No. DE-FG02-05ER46241 at MIT, and the U.S. DOE, Office of Science, Office of Basic Energy Sciences, under Contract No. DE-AC02-06CH11357 at Argonne National Laboratory.

\end{document}